\documentclass{ws-procs9x6}

\begin{document}
\vspace*{4cm}
\title{STATUS AND PERSPECTIVES OF DIRECT DARK MATTER SEARCHES}

\author{ G. CHARDIN }

\address{ DSM/DAPNIA/SPP, CEA/Saclay,\\
F-91191 Gif-sur-Yvette Cedex, France}

\maketitle\abstracts{
Supersymmetric particles represent the best motivated candidates
to fill the Dark Matter gap, and are actively hunted by a number
of competing experiments. Discriminating experiments are testing
for the first time SUSY models compatible with accelerator constraints.
These experiments contradict the 60 GeV WIMP candidate reported by the DAMA experiment.
The sensitivities of direct and indirect detection techniques for
both present experiments and future projects are compared.}

\section{Introduction : motivations}
The present situation of our knowledge of cosmological parameters is paradoxical. After the recent satellite MAP CMB measurements\cite{wmap}, the precision on the universe density is $\sim 1.02 \pm 0.02$, and the case for Dark Matter, which could still be considered arguable a few years ago, is now compelling. The total baryonic density, $\Omega_{baryon}$, is impressively constrained by primordial nucleosynthesis and cosmological constraints\cite{burles} to be approximately 4.5\%, implying that matter is composed at nearly 85\% of an as yet unobserved and mostly non interacting component, rather generically predicted by supersymmetric (SUSY) theories models. On the other hand, the recent apparition in the cosmological landscape of a non zero cosmological constant or some other quintessential component has brought some uneasiness to this Standard Cosmological Model: our Universe appears to be a strange mixture of $2/3$ of some cosmological repulsive component, $1/3$ of exotic matter, with only a few percent of ordinary matter. Worse, although CDM appears essential to produce cosmic structures observed at our present epoch, agreement with observations is marginal without additional components, such as neutrinos.

On the positive side, for the first time, direct detection experiments are beginning to test regions of supersymmetric model parameter space compatible with cosmological and accelerator constraints. We refer the reader to the review by Bergstrom for the Dark Matter phenomenology\cite{bergstrom} and will summarize the important effort undertaken by several groups, in both direct and indirect searches, to test a larger, if possible exhaustive, sample of SUSY parameter space.

\section{ WIMP direct detection : initial results and the DAMA candidate }

Initial direct detection experiments used detectors dedicated to other purposes, e.g. double--beta decay search, using conventional germanium detectors\cite{heidelberg94}, or sodium iodide NaI scintillating crystals\cite{dama96,ukdmc96,gerbier99}. In a first series of measurements, the Heidelberg--Moscow experiment, using a set of ultrapure isotopically enriched Ge crystals, established that massive neutrinos could not represent the solution to Dark Matter over essentially all the cosmologically relevant mass interval\cite{heidelberg94}. Further improvements of the sensitivity of this experiment were mostly due to the passive reduction of internal $^{68}$Ge cosmogenic activation by deep--underground storage\cite{heidelberg99}. Attempt to use an anti--Compton strategy resulted in the HDMS well--type germanium detector\cite{hdms01} which, although efficient at MeV energy, resulted in only a factor two gain at the low energies (a few keV) relevant for WIMP searches. The IGEX experiment\cite{igex} is reaching a better sensivitivy over most of the WIMP mass range but remains beyond the sensitivity required to test the first SUSY models.

On the other hand, massive sodium iodide crystals have been used, notably by the DAMA, the UKDMC and the Saclay groups, to reach sensitivities of the order of 10$^{-5}$ picobarn. Despite the NaI inefficient discrimination at low energies, where the number of collected photons is small and the scintillation time constants are less separated, the DAMA experiment, using a total mass of $\sim 100$ kg of high purity NaI crystals, has reported in 1998 a first indication of an annual modulation\cite{dama9899} using a data set of $\sim$12.5 kg $\times$ year, recorded over a fraction of a year. Apart from the ELEGANT--V experiment\cite{ejiri}, which is using NaI scintillators of total mass 730 kg, the DAMA experiment is presently running the largest experiment for WIMP direct detection. Compared to ELEGANT--V, DAMA is using NaI crystals with a lower radioactive background, with differential rates at low energies of $\sim$ 2-3 events/kg/keV/day down to an energy of 2 keV electron equivalent (e.e.), or $\sim 25$ keV recoil energy.

\begin{figure}[bht]
\centerline{\epsfxsize=3.9in\epsfbox{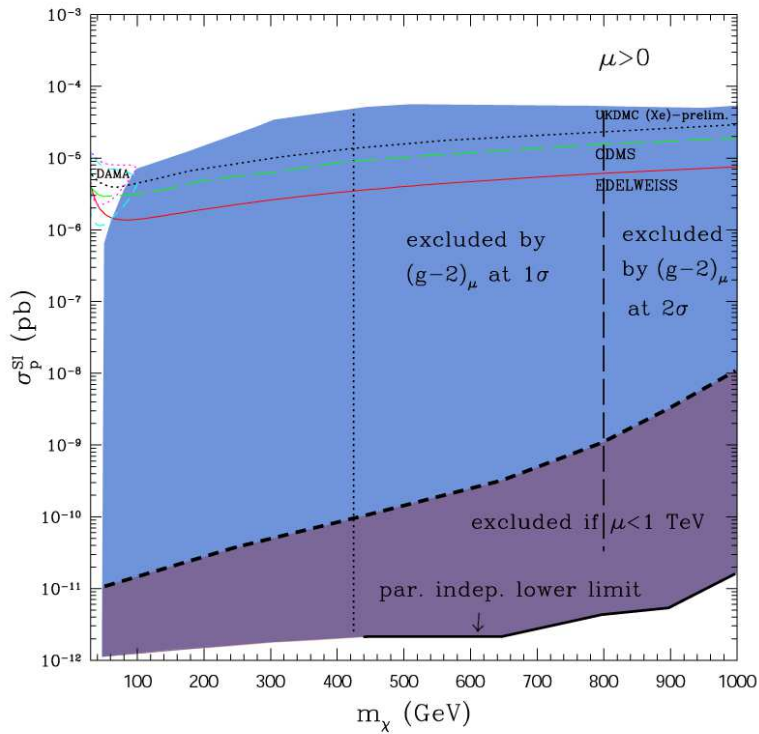}}   
\caption{ Experimental sensitivities of the present most sensitive WIMP direct detection experiments (from Ref.\protect\cite{kim} ). The EDELWEISS result, without background, now excludes the full 3-$\sigma$ zone of the DAMA signal compatible with accelerator constraints, independently of the WIMP halo model parameters. \label{roszkowski}}
\end{figure}

After confirming the annual modulation using a second data set of $\sim$ 41 kg $\times$ year\cite{dama9899}, the DAMA group published in 2000 an analysis involving a 160 kg $\times$ year data sample recorded over a three year time interval\cite{dama00}. Taken at face value, the DAMA observation presents a $4.5\sigma$ statistical significance, with both phase and amplitude consistent over a period of three years with a WIMP signature. Interpreted in terms of a WIMP candidate, the mass appears to be $\sim (52 \pm 10)$ GeV and the WIMP--nucleon cross--section $\sim (7 \pm 1) 10^{-6}$ picobarn. The allowed region, delimited by a three sigma contour, is represented in Fig.~\ref{roszkowski} together with the constraints of the presently most sensitive experiments.

\section{ WIMP direct detection : discriminating experiments }

Much of the progress of recent direct detection experiments is related to background discrimination capabilities of a new generation of detectors. Three main techniques have been developed successfully over the last ten years. Cryogenic experiments, EDELWEISS\cite{stefano}, CDMS\cite{shutt92}, CRESST\cite{bravin} and ROSEBUD\cite{cebrian02}, have built detectors capable of the simultaneous detection of two signals: ionisation and phonon signals for CDMS and EDELWEISS, scintillation and phonon signals for the CRESST and ROSEBUD experiments.

\begin{figure}[bht]
\centerline{\epsfxsize=3.9in\epsfbox{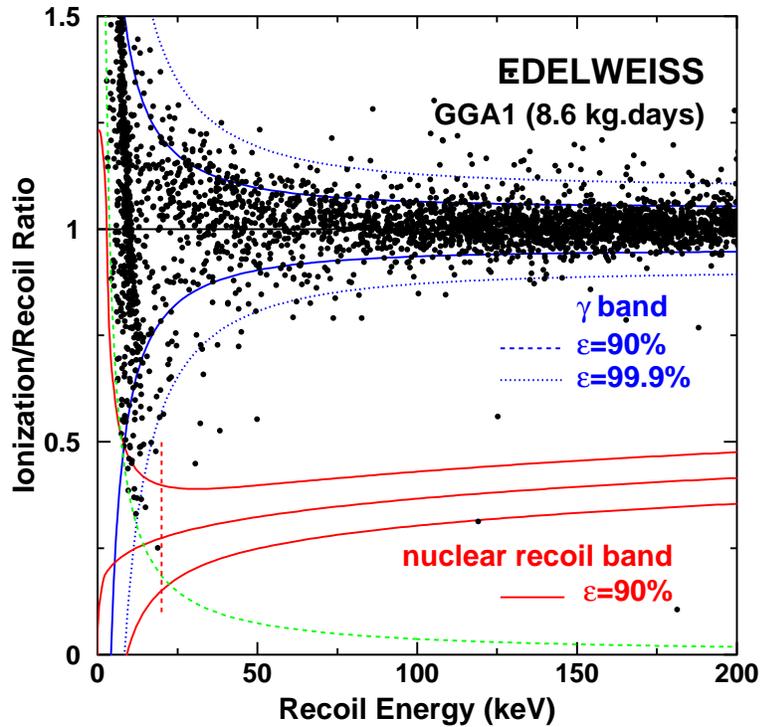}}   
\caption{ Scatter diagram of the ionisation efficiency, normalized to electron recoils, as a function of recoil energy for all events with energy $< 200$ keV recorded by the EDELWEISS experiment in the fiducial volume of a 320 gram Ge detector (from Ref.\protect\cite{edelw02}). With an effective mass 600 smaller than the DAMA NaI crystals, and an exposure 10 000 times shorter, this detector exceeds by a factor $> 5$ the sensitivity of the DAMA experiment. \label{edw_scatter_plot}}
\end{figure}

In 2000, the CDMS experiment\cite{cdms00,cdms02}, set in the shallow Stanford Underground Facility, excluded a large fraction of the 3-$\sigma$ DAMA zone. However, CDMS suffered from a significant neutron background (27 events were observed for a 15.8 kg $\times$ day exposure). On the other hand, EDELWEISS\cite{edelw01,edelw02}, in two background free data takings with a total exposure of 12 kg $\times$ day (Fig.~\ref{edw_scatter_plot}), clearly excludes the whole DAMA region compatible with accelerator constraints. The DAMA group has contested this contradiction, invoking the uncertainty in the WIMP halo parameters\cite{dama02}. But Copi and Krauss\cite{copi} have recently shown that the contradiction is model--independent when the relative sensitivity of both experiments is considered.
Therefore, unless unconventional WIMP--nucleon couplings are used, the DAMA candidate must now be considered excluded.

Using a background discrimination based on the different scintillation time constants for nuclear and electron recoils, the ZEPLIN collaboration\cite{zeplin} has recently obtained a promising result. In a 90 kg $\times$ day data sample using a 4.5 kg liquid cell, ZEPLIN is announcing a sensitivity within a factor 2 of that of EDELWEISS. However, the electronic background rate, probably due to an internal krypton contamination, is 50 times higher than the CDMS or EDELWEISS $\gamma$-ray background rate. Also, the energy resolution is much poorer than that of the cryogenic detectors : at 40 keV nuclear recoil energy (8 keV electron equivalent), the energy resolution is $\sim 100\%$. Additionally, no calibration exists for nuclear recoils below 50 keV recoil energy, and there is a considerable discrepancy between the quenching factor measurements realized by the DAMA and by the ZEPLIN groups. These two parameters must be determined before the present ZEPLIN sensitivity can be considered as established.

\section{ WIMP direct detection : future projects }
The present EDELWEISS result\cite{edelw02} ---no nuclear recoil candidate event over a 3-month period with a fiducial detector mass of 180 gram--- corresponds to a WIMP--nucleon cross--section $\sigma \sim 10^{-6}$ picobarn. This gives an idea of the difficulty to reach the $10^{-8}$ picobarn or, for that matter, the $10^{-10}$ picobarn milestone required to sample, respectively, the more realistic SUSY models\cite{ellis00} or most of the SUSY parameter space\cite{kim}.
Two non--discriminating experiments, CUORE\cite{cuore} and GENIUS\cite{genius} are proposing to meet the challenge of direct detection at the level of $10^{-8}$ pbarn or below. But reaching this sensitivity will require three orders of magnitude improvement over the presently achieved background levels. Also, these experiments are unable, if they observe candidate events, to demonstrate that these are due to WIMP interactions, except through the challenging annual modulation technique.
On the other hand, CDMS, CRESST and EDELWEISS are presently upgrading to detector mass between 10 kg for CDMS and CRESST, to 35 kg for EDELWEISS. ZEPLIN will be moving to a two--phase (liquid--gas) operation allowing scintillation and ionisation to be measured simultaneously, with a xenon target mass of 30 kg. These four experiments all promise a target sensitivity of the order of 2 10$^{-8}$ pbarn, just at the level of the models considered as realistic by Ellis et al. \cite{ellis00}. Beyond these experiments, in Europe, in the US and in Japan, tonne--scale cryogenic and xenon detectors are considered with the GENIUS, CUORE, CryoArray, Majorana and XMASS projects. Clearly, the scientific impact of a detection will be much higher and more robust if complementary informations are recorded using at least two target nuclei.

\section{ WIMP direct and indirect detection : complementarity and compared sensitivities}
Despite their small interaction cross--section with ordinary matter, WIMPs can be captured by celestial bodies, such at the Sun or the Earth\cite{indirect}. Since neutralinos are massive Majorana particles, they can annihilate and release copious fluxes of neutrinos, giving rise to observable signals in large--size terrestrial detectors. Annihilation at the galactic center, in the vicinity of the massive black hole at the center of our Milky Way, has been also considered as a possible copious source of annihilations, but the uncertainties in the density enhancement factor makes its flux extremely imprecise.
The overwhelming muon background coming from the above horizon hemisphere imposes to have a detector with directional capabilities, to distinguish upward going muons, associated to neutrino interactions, from the down--going cosmic--ray remnants. Cerenkov detectors provide an elegant solution to this experimental challenge, with large and unexpensive target mass.
Present experiments\cite{amanda,antares,macro,baksan,superk} include Baksan, Macro, now dismantled, and SuperKamiokande for the deep underground detectors, and AMANDA and Baikal for, respectively, under--ice and underwater detectors.

The presently most sensitive experiment for spin--independent WIMP interactions, SuperKamiokande, using a 3.5 years data sample, has recently published\cite{superk} a sensitivity limit, based on the analysis of Kamionkovski et al. \cite{kamionkowski}, of the same order but somewhat less sensitive than the recent EDELWEISS result\cite{edelw02}. AMANDA--B and Baksan are reaching somewhat lesser but similar sensitivities, with a higher energy threshold for the former experiment.
Future experiments include ANTARES, a European collaboration in the Mediterranean sea, and ICECUBE, a km$^{2}$ extension of the second generation AMANDA--B detector.
ANTARES, in its 0.1 km$^{2}$ version, plans to increase the present indirect detection sensitivity by a factor $\sim 3$ and ICECUBE is expected to increase the ANTARES sensitivity by a further order of magnitude, at least for high WIMP mass. This experiment benefits from a larger detection area, in the km$^{2}$ range, but the diffusion of Cerenkov photons in the ice is expected to lead to a partially degraded angular resolution at low muon energies.
In conclusion, indirect detection experiments will hardly be competitive for spin--independent couplings but are complementary to the direct searches due to their better sensitivity for predominantly axial, or spin--dependent, couplings (e.g. pure gauginos).
\section{Conclusions}
WIMP direct detection experiments are finally reaching sensitivities allowing to sample SUSY models compatible with accelerator constraints. The first WIMP candidate proposed in 2000 by the DAMA experiment is now clearly excluded by the EDELWEISS result, without any background subtraction and independently of galactic WIMP models unless unconventional interaction models are used.
Over the next few years, a second generation of discriminating experiments, CDMS--II, EDELWEISS--II, CRESST--II and ZEPLIN--II, using mass targets in the 10-30 kg range, intend to reach the impressive sensitivity of $10^{-8}$ picobarn, allowing to test a much larger fraction of realistic SUSY models. Direct searches with a detector mass of the order of one ton should be able to test most of the SUSY parameter space. Reaching a sensitivity of $10^{-10}$ picobarn, however, will require outstanding background discrimination capabilities, as well as a control of the neutron background.
Indirect detection experiments, such as ICECUBE or ANTARES--II, being more sensitive to the spin--dependent part of the interaction, are complementary to direct detection experiments and may help determine the nature of a WIMP candidate.
Improvements in sensitivity by WIMP direct and indirect detection experiments will hopefully allow to detect and identify the nature of Dark Matter within the next few years.

\end{document}